\documentclass[10pt,twocolumn,a4paper]{article}

\usepackage[utf8]{inputenc}
\usepackage[margin=0.75in]{geometry}
\usepackage{microtype}
\usepackage{amsmath,amssymb}
\usepackage{graphicx}
\usepackage{xcolor}
\usepackage{booktabs}
\usepackage[font={small},labelfont=bf]{caption}
\usepackage{subcaption}
\usepackage{cite}
\usepackage{url}
\usepackage[colorlinks=true, linkcolor=black, citecolor=black, urlcolor=black]{hyperref}
\usepackage{orcidlink}

\graphicspath{{../figures/paper/}{figures/paper/}{figures/}{.}}

\providecommand{\doi}[1]{\href{https://doi.org/#1}{\nolinkurl{doi:#1}}}

\title{\textbf{\LARGE Finding Low-Energy States of the Bernasconi Model via Simulated Transverse-Field Quantum Dynamics}}

\author{
    \textbf{Shaashwat Sagar Kale}\,\orcidlink{0009-0008-5202-2937} \quad 
    \textbf{Rudranil Basu}\,\orcidlink{0000-0003-0655-0890}\\[0.3em]
    \small Department of Physics, BITS Pilani, K K Birla Goa Campus, Goa, India\\[-0.2em]
    \small \textit{Email: f20220206@goa.bits-pilani.ac.in, rudranilb@goa.bits-pilani.ac.in}
}

\date{}
\begin{document}

\maketitle

\begin{abstract}
The Low-Autocorrelation Binary Sequence (LABS) problem, the ground-state
problem of the Bernasconi model, has a rugged landscape with deep, isolated
local minima. We study real-time transverse-field dynamics initialized in
these minima as a means of reaching lower-energy configurations. Using GPU
statevector simulation for $N=20$--$30$ spins, we
compare the escape probability per verified quantum sample with the success
rate per energy evaluation of steepest-descent kicks, random restarts and
tabu search. Using the same field and sampling window across all
system sizes, we find that, relative to the best classical
baseline, the sampler achieves a factor of $11.9$--$35.2$ for the deepest
traps and $13.4$--$27.2$ for the next energy tier. These ratios describe a
constant-factor improvement in sample efficiency over the sizes studied.
At the adopted field, broad Hamming-space spreading coexists with a strong
preference for low-energy states. This regime follows a sharp increase
in displacement as the field is raised. To connect this behaviour to the
landscape, a short-time expansion relates the onset
of escape to target distance and intermediate energies. Bounds on isotropic
spin flips and on the free transverse mixer quantify the role of the
Bernasconi potential in the observed transfer. At $N=28$, the
ground-state probability at $t=30$ exceeds a canonical cost prediction
fitted to the same mean energy by a factor of about $1.9$.
The cost of generating each
quantum sample is assessed separately through a hardware estimate, which
places a direct circuit implementation two to three orders of magnitude
beyond present processor capabilities.

\end{abstract}
\section{Introduction}
\label{sec:intro}
We study whether coherent spin dynamics can improve the escape from a
deep classical minimum. The Bernasconi model provides a concrete setting
in which to connect this question to optimisation of binary sequences and to
earlier work on quantum search.

Binary sequences with small aperiodic autocorrelation are handy in radar
ranging, pulse compression and spread-spectrum communication. Their design
leads to the Low-Autocorrelation Binary Sequence (LABS) problem, whose
quality measure is the merit factor~\cite{golay1977,golay1982,jedwab2005}.
Constructive families and correlation bounds have clarified the attainable
merit factors~\cite{jedwab2013,schmidt2016}. Even with this progress, finding
the optimal sequence at a given length remains a demanding discrete search
problem.

Bernasconi formulated this search as the ground-state problem of a
long-range Ising model~\cite{bernasconi1987}. Its deterministic interactions
produce a rugged energy landscape with many local minima, making it a
practical setting for studying glassy behaviour without quenched
disorder~\cite{bouchaud1994,marinari1994,mertens1998}. The distribution and structure of
these minima help explain the difficulty of local search~\cite{ferreira2000}.
To assess search methods on this landscape, exhaustive and branch-and-bound
calculations provide exact ground states and reliable benchmarks at finite
sequence lengths, although their computational cost grows exponentially
with length~\cite{mertens1996,packebusch2016}.

Classical methods address this difficulty through structured local search,
restarts and combinations of global and local moves. For example, relaxation-based
search, memetic algorithms and self-avoiding walks have produced strong
LABS solvers~\cite{prestwich2007,gallardo2009,boskovic2017,boskovic2024}.
The inclusion of LABS in the Quantum Optimization Benchmarking Library
also makes it a common setting for comparing quantum and classical
optimization~\cite{qoblib2026}. Within this broader search problem,
escape from a local minimum is a basic step shared by many heuristics. Iterated local search and basin hopping
perturb the current configuration and then descend again, while multistart
methods explore fresh basins~\cite{lourenco2003,wales1997,marti2003}.
The difficulty of this step becomes clear from the scarcity of other beneficial
proposals. For the deepest traps studied here, the fraction of lower-energy
configurations falls from $7.6\times10^{-6}$ at $N=20$ to
$1.5\times10^{-8}$ at $N=30$ spins. This scarcity makes the distribution of
escape proposals central to the performance of local search.

Thermal and quantum fluctuations offer different ways to explore such
landscapes. Simulated annealing uses thermal acceptance of uphill
moves~\cite{kirkpatrick1983}, while quantum annealing introduces a kinetic
term that couples classical configurations~\cite{finnila1994,kadowaki1998}.
Adiabatic algorithms, quantum annealing studies and finite-range tunnelling
benchmarks establish the role of this coupling in optimization
dynamics~\cite{farhi2001,santoro2002,denchev2016}. The corresponding algorithms and implementations are discussed in
Refs.~\cite{albash2018,hauke2020}. A recurring physical issue in these studies is
localization in configuration space,
including its effect on adiabatic optimization~\cite{altshuler2010}.

With the Hamiltonian held fixed, the same coupling defines a
continuous-time quantum walk on the graph of spin configurations.
This viewpoint brings the escape problem into the framework of coherent
transport and search~\cite{farhi1998,childs2004,mulken2011}, with the hypercube supplying
the natural graph for independent spin flips~\cite{moore2002}. In addition to that, quantum-walk
studies of spin-glass optimization connect the competition between the cost
Hamiltonian and the driver to the probability of finding low-energy
states~\cite{callison2019,callison2021}.

Our protocol is closely related to population transfer, in which evolution
starts from a low-energy classical configuration and redistributes
amplitude among configurations of comparable
energy~\cite{kechedzhi2018,smelyanskiy2020,baldwin2018}. Work on the quantum
random energy model relates this process to localization, mobility edges
and extended states~\cite{laumann2014,baldwin2016,faoro2019}. Numerical
studies further examine how the field and evolution time determine the
transfer~\cite{parolini2020}. The choice of a convenient initial classical
state also connects this approach to quantum local search and reverse
annealing~\cite{chancellor2017,ohkuwa2018}. Quantum-generated proposals
have been studied in Markov chain Monte Carlo~\cite{layden2023}, and
classical warm starts have been incorporated into variational quantum
optimization~\cite{egger2021}.

For LABS itself, the quantum approximate optimization algorithm
(QAOA)~\cite{farhi2014,zhou2020} has been compared with strong classical
solvers, with evidence of a scaling advantage reported for a
fixed-parameter QAOA-based approach~\cite{shaydulin2024}. Along a different route, a recent
variational study uses Pauli correlation encoding to represent larger
LABS instances with few qubits~\cite{sciorilli2025}. These developments
motivate a complementary question about a single optimization move.
Given a deep classical minimum, how effectively can transverse-field
dynamics produce a configuration of lower energy?

We address this question directly on the deterministic Bernasconi
landscape. The initial states exhaust the symmetry-distinct minima in
the two lowest excited energy tiers. Full-statevector simulations for
$N\in\{20,22,24,26,28,30\}$ compare the resulting sample yield with
classical escape methods through their successes per energy evaluation. To understand the resulting
escape probabilities, we develop a short-time expansion, geometric bounds
on isotropic kicks and the free mixer, constraints from energy
conservation, and a comparison with a canonical cost distribution
fixed by the measured mean energy. These results connect the numerical observations to the
distance, number and energies of the accessible configurations.

\section{Model and escape problem}
\label{sec:model}

Our first step is to make the escape problem precise. We introduce the
classical energy and the transverse-field dynamics, and then identify the
initial traps and the lower-energy configurations that count as successful
escape. This gives a common setting for the quantum and classical searches.

\subsection{Energy and spin dynamics}
We represent a binary sequence by a configuration of $N$ spins,
$s=(s_1,\dots,s_N)$, with $s_i\in\{-1,+1\}$. Its overlap with a
shifted copy is measured by the aperiodic autocorrelation. For a lag
$k\in\{1,\dots,N-1\}$, this is
\begin{equation}
C_k(s) = \sum_{i=1}^{N-k} s_i s_{i+k}.
    \label{eq:autocorrelation}
\end{equation}
To favour sequences with small correlations at all nonzero lags, we
combine these overlaps into the classical Bernasconi energy,
\begin{equation}
E(s) = \sum_{k=1}^{N-1} C_k(s)^2.
    \label{eq:bernasconi_energy}
\end{equation}
The merit factor $F(s)=N^2/[2E(s)]$ measures sequence quality, so
maximizing $F(s)$ is equivalent to minimizing $E(s)$~\cite{golay1982,jedwab2005}.
Expanding Eq.~\eqref{eq:bernasconi_energy} gives the long-range spin
interactions~\cite{mertens1998,packebusch2016,marinari1994},
\begin{equation}
E(s) = \frac{N(N-1)}{2} + 2 \sum_{k=1}^{N-1} \sum_{1 \le i < j \le N-k} s_i s_{i+k} s_j s_{j+k}.
    \label{eq:expanded_energy}
\end{equation}

The energy is invariant under global spin inversion
($s_i\mapsto-s_i$), sequence reversal ($s_i\mapsto s_{N-i+1}$) and
alternating sign reversal ($s_i\mapsto(-1)^i s_i$). These operations
generate an order-$8$ symmetry group~\cite{packebusch2016}.
We therefore choose one canonical representative from each orbit when
identifying distinct initial traps.

Having specified the energy landscape, we introduce dynamics between
its configurations by replacing each spin with a Pauli-$Z$ operator
$\hat{Z}_i$. The resulting diagonal cost
Hamiltonian satisfies $\hat{H}_B|s\rangle=E(s)|s\rangle$, where
$|s\rangle$ is the computational-basis state representing the string $s$.
During evolution, this term assigns an energy-dependent phase to each
configuration, while a transverse field of strength $g$ produces the
transitions between them. The full Hamiltonian is therefore

\begin{equation}
    \hat{H}(g) = \frac{1}{N}\hat{H}_B - g \sum_{i=1}^N \hat{X}_i,
    \label{eq:transverse_hamiltonian}
\end{equation}
where $\hat{X}_i$ is the Pauli-$X$ operator acting on spin $i$, and the factor $1/N$ ensures that both terms scale extensively. The transverse operator $\sum_i \hat{X}_i$ couples states differing by a single spin flip, allowing probability amplitude to move across the Boolean hypercube.

Let $J$ denote the physical energy scale multiplying $\hat{H}_B/N$.
We express Hamiltonians in units of $J$ and measure evolution time $t$
in units of $\hbar/J$, where $\hbar$ is the reduced Planck constant.
All quoted fields and times are numerical values in these units, and
we set $J=\hbar=1$ throughout.
\subsection{Traps and lower-energy states}
The escape problem begins at a configuration that resists a local
energy-lowering move. We call $s_0$ a local minimum, or trap, when every
single spin flip increases or preserves its energy,
\begin{equation}
E(s') \ge E(s_0) \quad \forall s' \text{ such that } d_H(s', s_0) = 1,
\end{equation}
where $d_H(s_a, s_b) = \sum_{i=1}^N \frac{1 - s_{a,i} s_{b,i}}{2}$ is the Hamming distance.

To quantify the difficulty of escaping a trap $s_0$ of energy $E_0 = E(s_0)$, we define the minimal barrier distance $d_{\min}$ as the minimum Hamming distance to any configuration of strictly lower energy,
\begin{equation}
d_{\min}(s_0) = \min_{s} \left\{ d_H(s, s_0) \;\middle|\; E(s) < E_0 \right\}.
    \label{eq:d_min}
\end{equation}
Distance alone gives only part of the target geometry. We must also
count how many lower-energy configurations lie at each distance. We denote the set of such configurations by
$\mathcal{S}_< = \{s : E(s) < E_0\}$. For an integer distance
$d \in \{0,\ldots,N\}$, let $M_d$ be the number of configurations in this
set at distance $d$ from $s_0$, and let $M$ be their total number.
Using vertical bars to denote the number of elements in a set,
\begin{equation}
M_d = \bigl|\{s\in\mathcal{S}_< : d_H(s,s_0)=d\}\bigr|,
    \qquad M = \sum_{d=0}^{N} M_d.
    \label{eq:target_counts}
\end{equation}
Uniform sampling of all configurations therefore has escape probability
$P_{\mathrm{blind}}=M/2^N$. These counts include every bitstring,
including symmetry partners, whereas the seed counts in
Table~\ref{tab:seed_levels} refer to symmetry-distinct orbits.

{}{A large $d_{\min}$ places the nearest successful
configurations many spin flips away. Within the Hamming shell at distance
$d$, which contains $\binom{N}{d}$ strings, the target fraction is
$M_d/\binom{N}{d}$, so escape must overcome both distance and scarcity.}

To include the deepest traps systematically, we enumerate all one-flip
local minima from the full table of $2^N$ energies and group them by energy
after symmetry reduction. Let $E^*$ denote the global minimum energy and
$E_0^*$ the lowest local-minimum energy strictly above it. We call the
symmetry-distinct minima at $E_0^*$ Level~1, while Level~2 consists of
those at the next local-minimum energy above $E_0^*$. This exhaustive
selection matters because random-start descent rarely reaches some of
these low-lying basins. For example, it reaches the tiers at $E=44$ for
$N=24$ and $E=53$ for $N=26$ in only $0.05\%$ and $0.30\%$ of runs,
respectively.

The benchmark covers $N\in\{20,22,24,26,28,30\}$, from
$2^{20}\approx1.05\times10^6$ to $2^{30}\approx1.07\times10^9$
configurations, where the lower-energy target set becomes increasingly
sparse. For comparison, a $2{,}000$-shot budget already covers more than $3\%$ of
the $2^{16}=65{,}536$ configurations at $N=16$, where blind random
sampling finds a ground state with $62\%$ probability.

Table~\ref{tab:seed_levels} gives the known global optimum
$E^*$~\cite{packebusch2016}, the two trap energies, their orbit counts and
the range of $d_{\min}$ for each size. The steepest-descent (SD) hit
fractions show how frequently random-start descent reaches these tiers.
With this classification in place, we evaluate every dynamical observable
and escape ratio below on the orbits in this table.

\begin{table}[t]
\centering
\small
\caption{Classification of exhaustive local minima tiers evaluated across system sizes $N \in \{20, 22, 24, 26, 28, 30\}$. For each $N$, the exact global optimum $E^*$~\cite{packebusch2016} is shown. Level 1 comprises all symmetry-distinct local minima at the lowest energy $E_0^*$ strictly above $E^*$. Level 2 comprises all orbits at the next energy level above $E_0^*$. SD hit fraction is the fraction of $10^{5}$ random-start steepest descents that end in any orbit of that tier.}
\label{tab:seed_levels}
\resizebox{\columnwidth}{!}{%
\begin{tabular}{ccccccc}
\toprule
$N$ & $E^*$ & Tier & $E_0$ & Orbits & $d_{\min}$ range & SD hit fraction \\
\midrule
20 & 26 & Level 1 & 34 &   4 & 6--8  & 1.03\% \\
   &    & Level 2 & 38 &   7 & 5--7  & 2.13\% \\
\midrule
22 & 39 & Level 1 & 47 &  29 & 4--9  & 3.10\% \\
   &    & Level 2 & 55 & 101 & 2--6  & 9.79\% \\
\midrule
24 & 36 & Level 1 & 44 &   1 & 8     & 0.05\% \\
   &    & Level 2 & 48 &   1 & 9     & 0.07\% \\
\midrule
26 & 45 & Level 1 & 53 &   6 & 6--9  & 0.30\% \\
   &    & Level 2 & 61 &  20 & 4--9  & 0.69\% \\
\midrule
28 & 50 & Level 1 & 58 &   1 & 10    & 0.05\% \\
   &    & Level 2 & 62 &   2 & 11    & 0.03\% \\
\midrule
30 & 59 & Level 1 & 67 &   4 & 10--13 & 0.05\% \\
   &    & Level 2 & 75 &  19 & 4--11 & 0.19\% \\
\bottomrule
\end{tabular}%
}
\end{table}

\section{Simulation and comparison}
\label{sec:methodology}

We now turn the escape problem into a common sampling test for quantum
dynamics and classical search. To do so, we specify the evolution and its
observables, and compare the classical success rates through their energy
evaluation cost. We then estimate the circuit resources needed to generate
the quantum samples.

\subsection{Time evolution and measured quantities}
To follow coherent evolution under Eq.~\eqref{eq:transverse_hamiltonian}, we simulate the full statevector $|\psi(t)\rangle \in \mathbb{C}^{2^N}$ using NVIDIA cuStateVec~\cite{custatevec} accelerated by custom CUDA kernels. We initialize the state in the computational-basis configuration of the trap, $|\psi(0)\rangle=|s_0\rangle$.

Write the diagonal part as $\hat{H}_Z=\hat{H}_B/N$ and the transverse
part as $\hat{H}_X=-g\sum_i\hat{X}_i$. For a time step $\delta t$, define
the symmetric second-order product formula~\cite{trotter1959,strang1968,suzuki1976,childs2021},
\begin{equation}
    \hat{U}_2(\delta t) = e^{-i \frac{\delta t}{2} \hat{H}_Z} e^{-i \delta t \hat{H}_X} e^{-i \frac{\delta t}{2} \hat{H}_Z},
    \label{eq:symmetric_step}
\end{equation}
which approximates $e^{-i\delta t\hat{H}(g)}$ with a local error of $\mathcal{O}(\delta t^3)$.

The implementation applies a full diagonal step followed by a transverse
step. Denote this update by
$\hat{U}_{\mathrm{code}}(\delta t)=e^{-i\delta t\hat{H}_X}e^{-i\delta t\hat{H}_Z}$.
Writing $\hat{D}=e^{-i\delta t\hat{H}_Z/2}$ for a half diagonal step gives
$\hat{U}_{\mathrm{code}}=\hat{D}^{-1}\hat{U}_2\hat{D}$.
For an integer number $m$ of steps, this relation becomes
$\hat{U}_{\mathrm{code}}^m=\hat{D}^{-1}\hat{U}_2^m\hat{D}$.
The initial basis state acquires only a common phase under $\hat{D}$,
while the final $\hat{D}^{-1}$ changes only the phases of the evolved
state's basis amplitudes. Consequently, the computational-basis probabilities agree
exactly with those from Eq.~\eqref{eq:symmetric_step} at every recorded
time. The escape probabilities, Hamming displacements and energy
distributions therefore inherit the symmetric formula's accuracy.

The full diagonal step applies $\exp(-i\delta t E(s)/N)$ in parallel
across all $2^N$ states using a custom kernel with hardware population
counts (\verb|__popcll|). The transverse mixer factorizes into single-qubit
rotations, with $\hat{I}$ denoting the identity operator,
\begin{equation}
e^{i \delta t g \sum_i \hat{X}_i} = \prod_{i=1}^N \left[ \cos(g \delta t) \hat{I} + i \sin(g \delta t) \hat{X}_i \right],
\label{eq:mixer_factorization}
\end{equation}
applied sequentially via \texttt{custatevecApplyMatrix}. Simulations employ a Trotter step $\delta t = 0.05$ across a total horizon $t \in [0, 60]$ ($1{,}200$ Trotter steps).

Once the state has evolved, we use its amplitudes
$c_s(t)=\langle s|\psi(t)\rangle$ to follow both escape and spatial
spreading. Summing their
squared magnitudes over the target set $\mathcal{S}_<$ defined in
Sec.~\ref{sec:model} gives the escape probability $P_<(t)$. We evaluate
its average $\langle P_<\rangle$ over a fixed time interval
$[t_{\mathrm{start}},t_{\max}]=[10.0,38.5]$. At the step
$\delta t=0.05$, this interval contains $571$ recorded times
$t_j=10+0.05j$, with $j=0,\ldots,570$. The reported average assigns
equal weight to these times,
\begin{align}
P_<(t) &= \sum_{s : E(s) < E_0} |c_s(t)|^2,
    \label{eq:escape_probability}\\
    \langle P_< \rangle &= \frac{1}{571}\sum_{j=0}^{570} P_<(t_j).
    \label{eq:window_average}
\end{align}
The interval begins after the initial delay in escape and provides a
common sampling protocol for every seed and system size. Equivalently,
$\langle P_<\rangle$ is the success probability when the measurement time
is chosen uniformly from the $571$ recorded times in this interval.

The window average measures the yield of the move, while the approach
to the first peak reveals how quickly that yield develops. Let
$t_{\mathrm{peak},1}$ denote the time of the first local maximum of
$P_<(t)$, using the overall peak when the trajectory rises monotonically.
The first-peak growth indicator is
\begin{equation}
    \left. \frac{dP_<}{dt} \right|_{\mathrm{rise}} = \frac{P_<(t_{\mathrm{peak}, 1})}{t_{\mathrm{peak}, 1}}.
\end{equation}
The figures label this quantity the \textit{initial rise rate}. It
measures growth up to a peak whose time depends on $g$.
Section~\ref{subsec:escape_geometry} relates this finite-time indicator to
the short-time onset of escape.

To relate this growth to motion through configuration space, we also
measure the mean Hamming displacement from the initial trap,
\begin{equation}
    \langle d_H(t) \rangle = \sum_{s} |c_s(t)|^2 \, d_H(s, s_0).
\end{equation}
Together, the displacement and escape probability distinguish spatial
spreading from the accumulation of weight on lower-energy configurations.

\subsection{Classical search and success rates}
We compare quantum escape with the perturbation and local-search moves
used in classical heuristics. Kick followed by descent is a standard
component of iterated local search and basin
hopping~\cite{lourenco2003,wales1997}. Tabu search~\cite{glover1997}
complements these moves by keeping a short memory of the recent search,
and underlies strong LABS algorithms~\cite{gallardo2009,boskovic2017}.
All baseline implementations use cached autocorrelation updates with
$\mathcal{O}(N)$ cost per proposed spin flip. Counting cost-function
evaluations supplies a common measure of sample yield, while the quantum
evolution resources are assessed in Sec.~\ref{hard_res_con}. Explicit
resource accounting is also central to broader optimization
benchmarks~\cite{qoblib2026}.

In kick plus steepest descent (Kick-SD), we flip $d$ randomly chosen
spins of $s_0$ and descend to a local minimum. For each
$d\in\{2,4,\ldots,12\}$, we run $50{,}000$ trials. The success rate
per evaluation is denoted by $r_{\mathrm{SD}}(d)$, with one evaluation
charged for the sampled string and $N$ for each descent step. The best
kick rate is $r^*_{\mathrm{SD}}=\max_d r_{\mathrm{SD}}(d)$.
Restart plus steepest descent (Restart-SD) explores
fresh basins by beginning from a uniformly random string. It uses
$50{,}000$ trials to estimate the rate
$r_{\mathrm{restart}}$ under the same accounting~\cite{marti2003}.

To allow the local search to move beyond repeated visits to the same
configurations, we also use tabu search. It begins either directly at
$s_0$ (Tabu-direct), after a
$d$-spin kick (Tabu-kick, optimized over $d$), or at a uniformly random
string (Tabu-restart). At each move, the tenure is chosen uniformly from
$[\max(2,\lfloor N/6\rfloor),\max(3,\lfloor N/3\rfloor)]$.
An aspiration criterion accepts a tabu move when it strictly improves
the best energy found. Each trial scans all $N$ single-flip neighbours
per move, charging one evaluation for the initial string and $N$ for
each scan, and stops upon reaching $E<E_0$ or after $100$ moves.
We run $50{,}000$ trials per mode and per tested kick distance and
denote the best rate by $r^*_{\mathrm{tabu}}$.

Each classical rate is the number of successful trials divided by the
total number of energy evaluations across all trials. The trial count is
fixed, while the total evaluation count depends on the search path and
stopping time. This accounting compares the methods through their yield
per energy evaluation.

We use the largest of these classical rates, denoted by
$r^*_{\mathrm{classical}}$, to compare with the quantum yield. The
quantum-to-classical escape ratio $R$ is therefore

\begin{equation}
R = \frac{\langle P_< \rangle}{r_{\mathrm{classical}}^{*}}, \qquad
r_{\mathrm{classical}}^{*} = \max\!\left(r^{*}_{\mathrm{SD}},\,
r_{\mathrm{restart}},\, r^{*}_{\mathrm{tabu}}\right).
    \label{eq:ratio_r}
\end{equation}
Thus $R$ measures the yield of lower-energy configurations per energy
evaluation, with one evaluation assigned to verification of each quantum
sample. State preparation and evolution determine the additional cost
of generating that sample. Tabu search is the stronger classical
baseline on most traps, exceeding the best steepest-descent variant by
a factor of $1.2$-$2$, with larger gains on shallow traps that
Tabu-direct escapes from the initial state.

Classical rates are estimated from the successes accumulated in the
$50{,}000$ trials. For a run with zero successes, let
$N_{\mathrm{eval}}$ be its total number of evaluations. We use
$1/N_{\mathrm{eval}}$ as the one-hit resolution of the rate, and mark
the corresponding $R$ as a lower-limit estimate under this convention.
For tiers with several orbits, we report medians and $68\%$ bootstrap
intervals using $B=2000$ resamples of the symmetry-distinct
orbits~\cite{efron1994}. A tier containing a single orbit is reported
as a single value, with orbit count $n=1$.

The target counts also allow a direct geometric comparison with the
perturbation itself. Consider a kick that
first chooses a distance $d$ with probability $w_d$ and then chooses uniformly
among the $\binom{N}{d}$ configurations at that distance from $s_0$.
Here $w_d\geq0$ and $\sum_{d=0}^{N}w_d=1$. We call this an isotropic kick and
denote its probability of reaching $\mathcal{S}_<$ by $P_{\mathrm{iso}}$.
Using the counts in Eq.~\eqref{eq:target_counts},
\begin{equation}
P_{\mathrm{iso}} = \sum_{d=0}^{N}w_d\frac{M_d}{\binom{N}{d}}
    \leq \max_{0\leq d\leq N}\frac{M_d}{\binom{N}{d}}.
    \label{eq:isotropic_shell_probability}
\end{equation}
Global spin inversion preserves the Bernasconi energy, so every target at distance $d$ has a target partner at distance $N-d$. Thus $M_d=M_{N-d}$, and targets occur only for $d_{\min}\leq d\leq N-d_{\min}$, with $d_{\min}$ defined in Eq.~\eqref{eq:d_min}. Throughout this interval, $\binom{N}{d}\geq\binom{N}{d_{\min}}$. Denoting the resulting upper bound by $B_{\mathrm{iso}}$, we obtain
\begin{equation}
P_{\mathrm{iso}}\leq B_{\mathrm{iso}}
    \equiv \frac{M}{\binom{N}{d_{\min}}}.
    \label{eq:isotropic_bound}
\end{equation}
Because the bound applies to every distribution of kick distances, it
also covers a radius chosen separately for each seed and arbitrary
mixtures of radii. It concerns the probability of reaching a target in
the kick itself. Subsequent descent and tabu moves use the energies of
visited configurations, so their performance is compared through
Eq.~\eqref{eq:ratio_r}. Together, these comparisons allow us to ask how
much of the quantum yield comes from reaching distant shells and how
much comes from favouring useful configurations within them.

\subsection{Circuit cost}
\label{hard_res_con}
The sample yield becomes practically beneficial on a processor once the required
evolution can be implemented. For the Bernasconi Hamiltonian, the main
circuit cost comes from its two- and four-body phases. Expanding $\hat{H}_B$ in Eq.~\eqref{eq:expanded_energy} yields $N(N-1)(N-2)/6$ four-spin products $s_i s_{i+k} s_j s_{j+k}$ before simplification. Products with $j = i + k$ reduce to two-body terms ($s_{i+k}^2 = 1$), and coincident index sets merge, leaving a mixture of two- and four-body Pauli-$Z$ operators. At $N = 30$, the Hamiltonian decomposes into exactly $210$ two-body and $1{,}925$ four-body commuting terms (and $182$ two-body and $1{,}547$ four-body terms at $N = 28$). Synthesizing these terms on an idealized all-to-all topology requires $2$ CNOTs per two-body and $6$ CNOTs per four-body term, totaling $\approx 12{,}000$ two-qubit gates per Trotter step naively ($\approx 9{,}650$ at $N = 28$).

\begin{table}[t]
\centering
\small
\caption{Two-qubit gate requirements for a single second-order Trotter step at $N = 28$ across physical hardware topologies compared to naive all-to-all synthesis.}
\label{tab:hardware_transpilation}
\resizebox{\columnwidth}{!}{%
\begin{tabular}{lcc}
\toprule
Architecture & Coupling Topology & 2Q Gates / Step \\
\midrule
Naive All-to-All & Idealized (unconstrained) & $\approx 9.6 \times 10^3$ \\
Square lattice (e.g.\ IBM Nighthawk) & degree-4 planar grid & $\approx 1.9 \times 10^4$ \\
IBM Heron & Heavy-hex planar & $\approx 3.0 \times 10^4$ \\
\bottomrule
\end{tabular}%
}
\end{table}

Routing on a processor with constrained connectivity increases the gate
count, as shown for $N=28$ in Table~\ref{tab:hardware_transpilation}.
To assess how time discretization affects the escape probability and establish
the largest step tolerable for hardware synthesis, Table~\ref{tab:trotter_check}
compares the window-averaged probability $\langle P_<\rangle_{[10,38.5]}$ for coarse
second-order Trotter steps with the reference at $\delta t=0.05$, using the deepest
traps at $N=16$, $20$ and $24$ ($g=0.35$). A step of $\delta t=0.25$ retains
$92$--$98\%$ of the reference probability (a loss of only $2$--$8\%$).
At $\delta t=0.5$, the loss grows to $31$--$84\%$, and steps of $\delta t\geq0.75$
retain at most $12\%$ of the escape probability. The rapid deterioration past
$\delta t\approx0.5$ matches the threshold $\delta t\lesssim0.5$ found for
non-commuting Hamiltonians by commutator-based error analysis~\cite{childs2021}.

\begin{table}[ht]
\centering
\small
\caption{Ratio of the window-averaged $\langle P_< \rangle$ at Trotter step
$\delta t$ to its value at $\delta t = 0.05$ (second-order formula, $g = 0.35$).}
\label{tab:trotter_check}
\begin{tabular}{lcccc}
\toprule
$N$ ($E_0$) & $\delta t = 0.25$ & $0.5$ & $0.75$ & $1.0$ \\
\midrule
16 (28) & 0.98 & 0.69 & 0.12 & 0.03 \\
20 (34) & 0.92 & 0.26 & 0.01 & 0.01 \\
24 (44) & 0.96 & 0.16 & 0.00 & 0.00 \\
\bottomrule
\end{tabular}
\end{table}

Using $\delta t = 0.25$ for the resource estimate, evolution to $t\in[10,30]$
requires $40$--$120$ steps. The resulting volume is of order $10^6$ two-qubit
gates, approximately $7\times10^5$--$2\times10^6$ on a square lattice and
$1\times10^6$--$4\times10^6$ on heavy-hex at $N=28$.

\section{Escape dynamics}
\label{sec:results}
With the sampling protocol fixed, we can ask which features of the
dynamics account for the enhanced escape. We first establish the gain
over classical search and relate it to the distance and energies of the
accessible states. We then examine how the transverse field and the
duration of evolution control the transfer into these states.

\subsection{Escape from deep minima}
\label{subsec:escape_geometry}
\begin{figure}[t]
    \centering
    \includegraphics[width=\columnwidth]{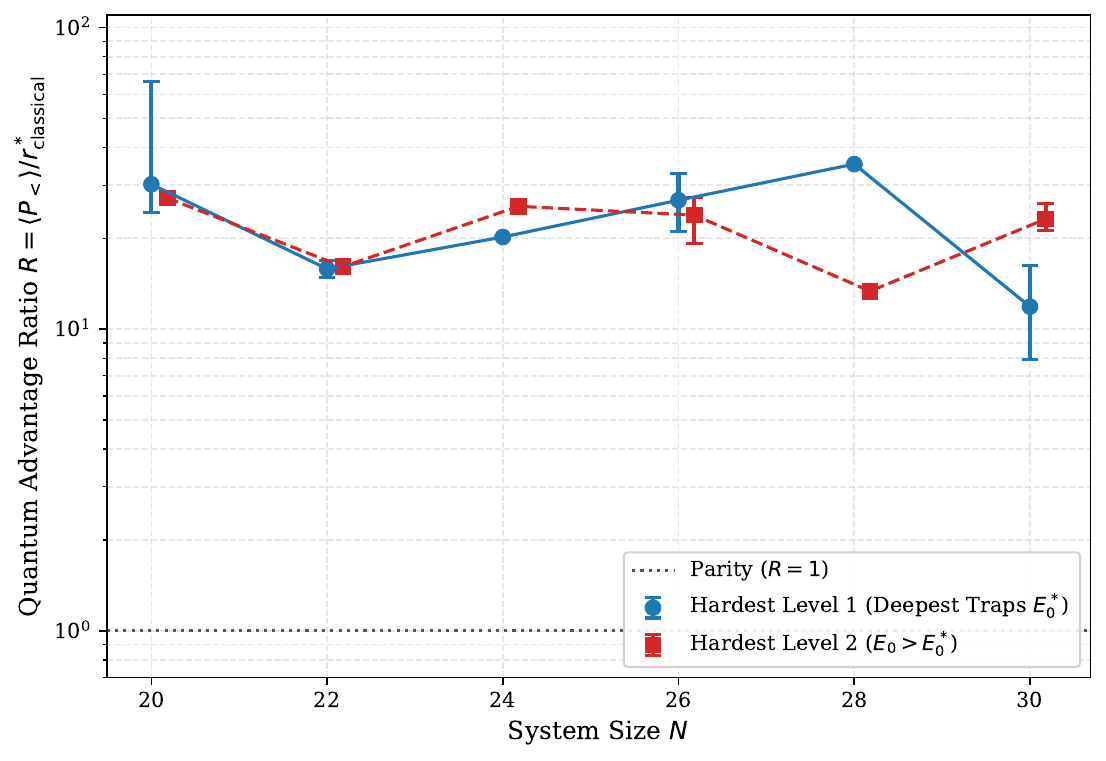}
    \caption{Quantum-to-classical escape ratio $R = \langle P_< \rangle / r_{\mathrm{classical}}^*$ versus system size $N$ for Level~1 (deepest traps) and Level~2 (second-hardest traps) at $g = 0.35$, against the best of steepest-descent kicks, restarts and tabu search. Points denote medians over symmetry-distinct orbits with $68\%$ bootstrap intervals ($B = 2000$). Upward arrows mark one-hit resolution estimates for zero-success classical runs. The dotted line denotes parity ($R = 1$).}
    \label{fig:ratio_r}
\end{figure}

We begin with the central comparison in Fig.~\ref{fig:ratio_r} and
Table~\ref{tab:ratio_summary}. They show the escape ratio
$R=\langle P_<\rangle/r_{\mathrm{classical}}^*$ for
$N\in\{20,22,24,26,28,30\}$, using Level~1 traps at $E_0^*$ and
Level~2 traps at $E_0>E_0^*$. The energy tiers, orbit counts and barrier
distances are those in Table~\ref{tab:seed_levels}. The same field
$g=0.35$ and sampling window $[10,38.5]$ are used at every size, including
$N=28$ and $30$, so the comparison follows one common escape protocol.

\begin{table*}[t]
\centering
\small
\caption{Quantum-to-classical escape ratio $R = \langle P_< \rangle / r_{\mathrm{classical}}^{*}$ at $g = 0.35$, window $[10, 38.5]$, where $r_{\mathrm{classical}}^{*}$ is the best of steepest-descent kicks, restarts and tabu search (Eq.~\eqref{eq:ratio_r}). Medians over symmetry-distinct orbits with $68\%$ bootstrap intervals. Entries marked $\ge$ use the one-hit resolution convention for zero-success classical runs. Levels follow Table~\ref{tab:seed_levels}. Orbits with shallow barriers $d_{\min} \le 2$ are marginal (a two-flip kick escapes them with $\approx 90\%$ probability).}
\label{tab:ratio_summary}
\resizebox{\textwidth}{!}{%
\begin{tabular}{lcccccc}
\toprule
\textbf{Energy Tier} & $N = 20$ & $N = 22$ & $N = 24$ & $N = 26$ & $N = 28$ & $N = 30$ \\
\midrule
\textbf{Level 1 (Deepest Traps)} & & & & & & \\
Trap Energy $E_0$ & 34 & 47 & 44 & 53 & 58 & 67 \\
Distinct Orbits ($n$) & 4 & 29 & 1 & 6 & 1 & 4 \\
Barrier Distance $d_{\min}$ Range & 6--8 & 4--9 & 8 & 6--9 & 10 & 10--13 \\
Sample Median $R$ & $\mathbf{30.22}$ & $\mathbf{15.84}$ & $\mathbf{20.20}$ & $\mathbf{26.75}$ & $\mathbf{35.24}$ & $\mathbf{11.87}$ \\
$68\%$ Bootstrap CI & $[24.28, 66.17]$ & $[14.87, 16.93]$ & $n=1$ & $[21.13, 32.79]$ & $n=1$ & $[7.93, 16.27]$ \\
\midrule
\textbf{Level 2 (Excited Traps)} & & & & & & \\
Trap Energy $E_0$ & 38 & 55 & 48 & 61 & 62 & 75 \\
Distinct Orbits ($n$) & 7 & 101 & 1 & 20 & 2 & 19 \\
Barrier Distance $d_{\min}$ Range & 5--7 & 2--6 & 9 & 4--9 & 11 & 4--11 \\
Sample Median $R$ & $\mathbf{27.22}$ & $\mathbf{16.10}$ & $\mathbf{25.53}$ & $\mathbf{23.87}$ & $\mathbf{13.36}$ & $\mathbf{23.18}$ \\
$68\%$ Bootstrap CI & $[26.26, 27.76]$ & $[15.41, 16.83]$ & $n=1$ & $[19.23, 27.25]$ & $[13.24, 13.48]$ & $[21.21, 26.04]$ \\
\bottomrule
\end{tabular}%
}
\end{table*}

The median ratio exceeds unity at every size in both tiers. More
specifically, Level~1 medians lie between
$11.9\times$ and $35.2\times$ and Level~2 medians between $13.4\times$ and
$27.2\times$ over $N = 20$--$30$ (Table~\ref{tab:ratio_summary}). The ratio
varies with the orbits present at each size and stays within a
constant-factor range over $N=20$--$30$. The same field and sampling
window give an enhanced yield also at $N=28$ and $30$.

\begin{figure*}[t]
\centering
\begin{subfigure}[b]{0.51\textwidth}
    \centering
    \includegraphics[width=\textwidth]{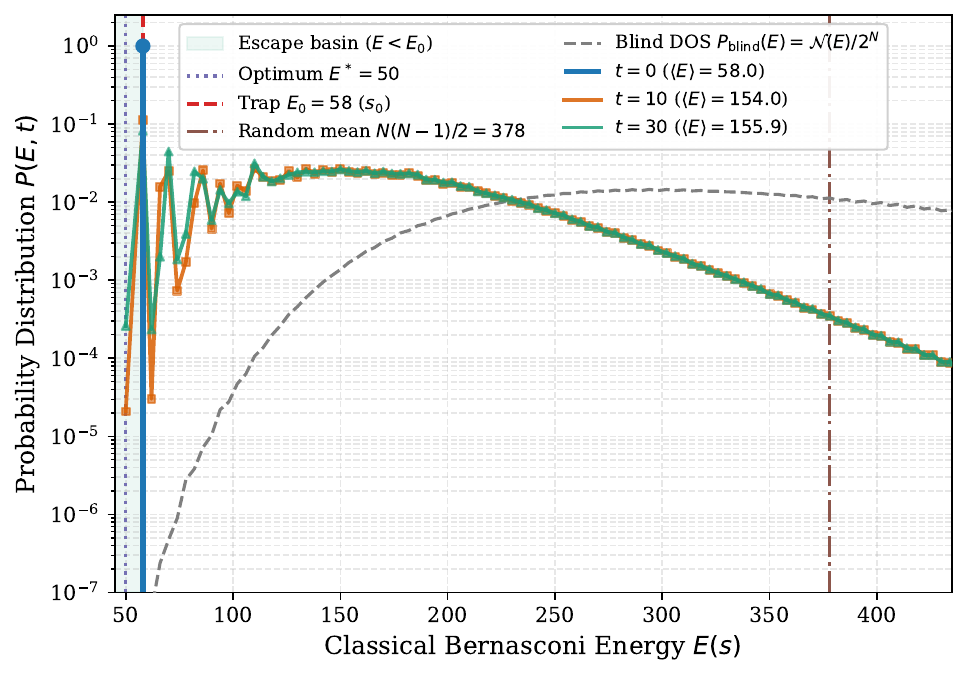}
    \caption{}
    \label{fig:energy_hist_a}
\end{subfigure}\hfill
\begin{subfigure}[b]{0.47\textwidth}
    \centering
    \includegraphics[width=\textwidth]{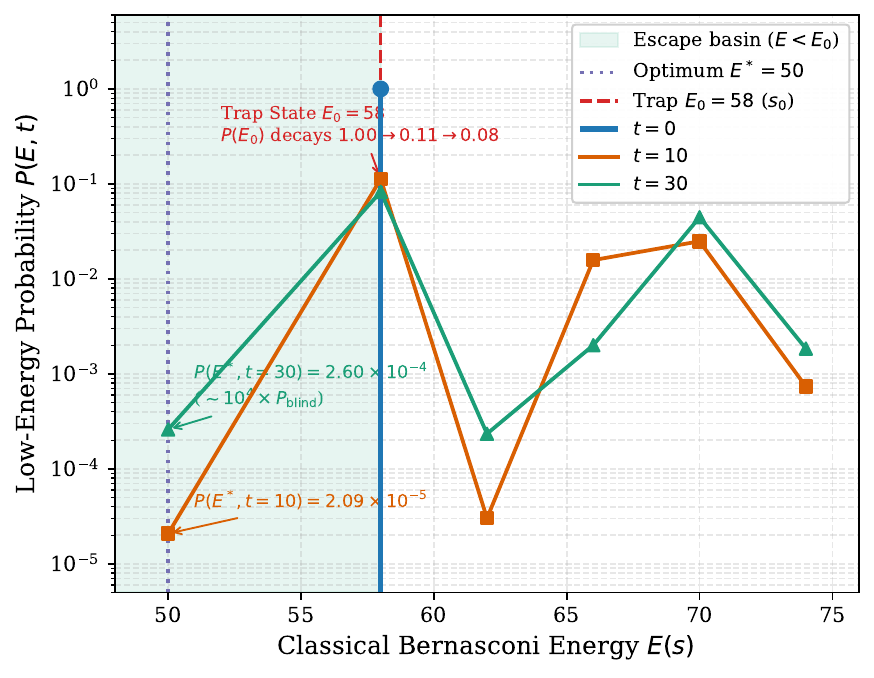}
    \caption{}
    \label{fig:energy_hist_b}
\end{subfigure}
\caption{Energy distribution of the quantum state $|\psi(t)|^2$ initialized in the deepest $N = 28$ trap ($E_0 = 58$, $g = 0.35$).
(a)~Full distribution at $t = 0, 10, 30$ compared to the blind density of states $P_{\mathrm{blind}}(E)=\mathcal{N}(E)/2^N$, where $\mathcal{N}(E)$ counts configurations of energy $E$ (grey dashed). While the mean energy $\langle E \rangle \approx 154$--$156$ remains far below the random-string expectation $N(N-1)/2 = 378$, the wavepacket broadens into low-lying states.
(b)~Detail of the low-energy region $E \le 76$. The ground-state level $E^* = 50$ ($8$ energy units below the trap) accumulates probability $2.60 \times 10^{-4}$ by $t = 30$, an enrichment factor of $\sim 10^4$ over blind sampling ($P_{\mathrm{blind}}(E^*) = 2.98 \times 10^{-8}$).}
\label{fig:energy_hist}
\end{figure*}

To understand this improvement, we turn to the distributions generated
by the escape moves. An isotropic kick
assigns equal probability to configurations at the chosen distance,
after which descent concentrates the samples in nearby minima. Quantum
evolution spreads amplitude over large Hamming distances
($\langle d_H\rangle\approx N/2$, Sec.~\ref{subsec:field_regimes})
while enhancing the weight of low-energy configurations relative to
blind sampling (Fig.~\ref{fig:energy_hist}). The coexistence of broad
spreading and enhanced low-energy weight leads us to examine the target
geometry and the phases accumulated along the paths between configurations.

\begin{figure*}[t]
    \centering
    \begin{subfigure}[b]{0.48\textwidth}
        \centering
        \includegraphics[width=\textwidth]{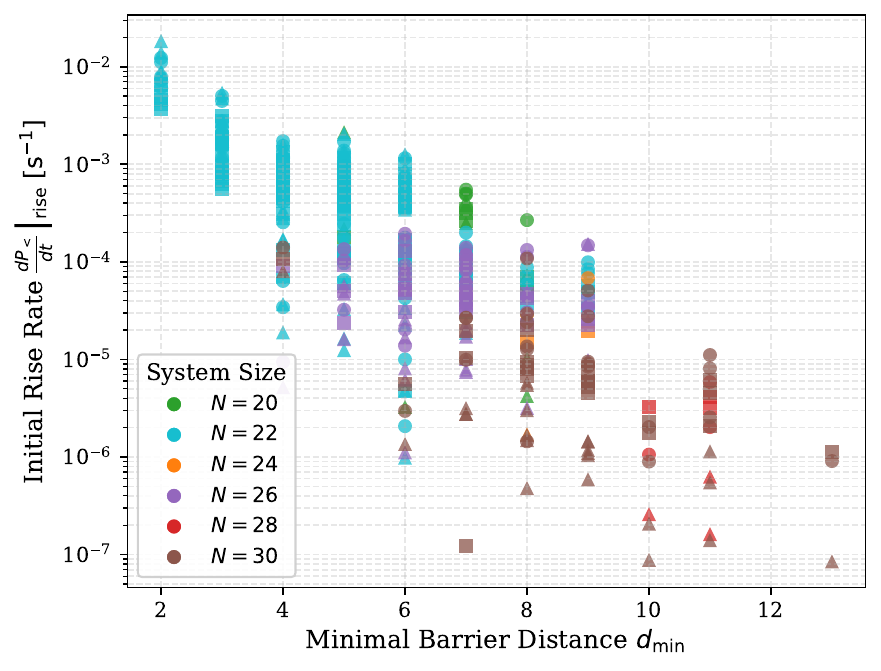}
        \caption{}
        \label{fig:rise_rate_dmin_a}
    \end{subfigure}\hfill
    \begin{subfigure}[b]{0.48\textwidth}
        \centering
        \includegraphics[width=\textwidth]{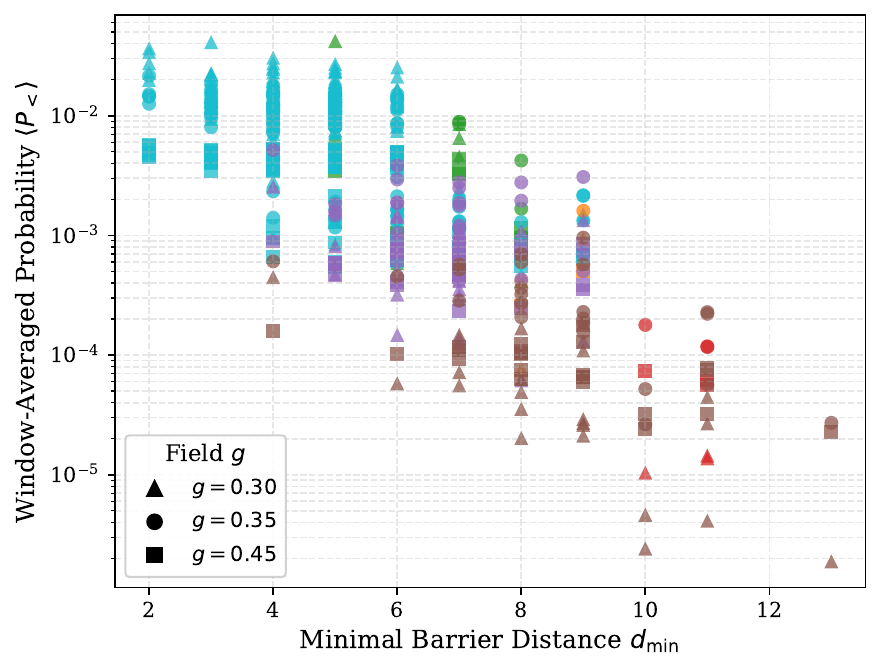}
        \caption{}
        \label{fig:rise_rate_dmin_b}
    \end{subfigure}
    \caption{Dependence of quantum escape dynamics on minimal barrier distance $d_{\min}$.
    (a)~Initial rise rate $\left.\frac{dP_<}{dt}\right|_{\mathrm{rise}}$ versus $d_{\min}$, showing the rapid decrease in the first-peak growth indicator as the distance increases.
    (b)~Window-averaged escape probability $\langle P_< \rangle$ versus $d_{\min}$, showing the range of later-time yields across target distances and energy tiers. Data points are colored by system size $N \in \{20, 22, 24, 26, 28, 30\}$ and shaped by coupling strength $g \in \{0.30, 0.35, 0.45\}$.}
    \label{fig:rise_rate_dmin}
\end{figure*}

\begin{figure*}[t]
    \centering
    \begin{subfigure}[b]{0.48\textwidth}
        \centering
        \includegraphics[width=\textwidth]{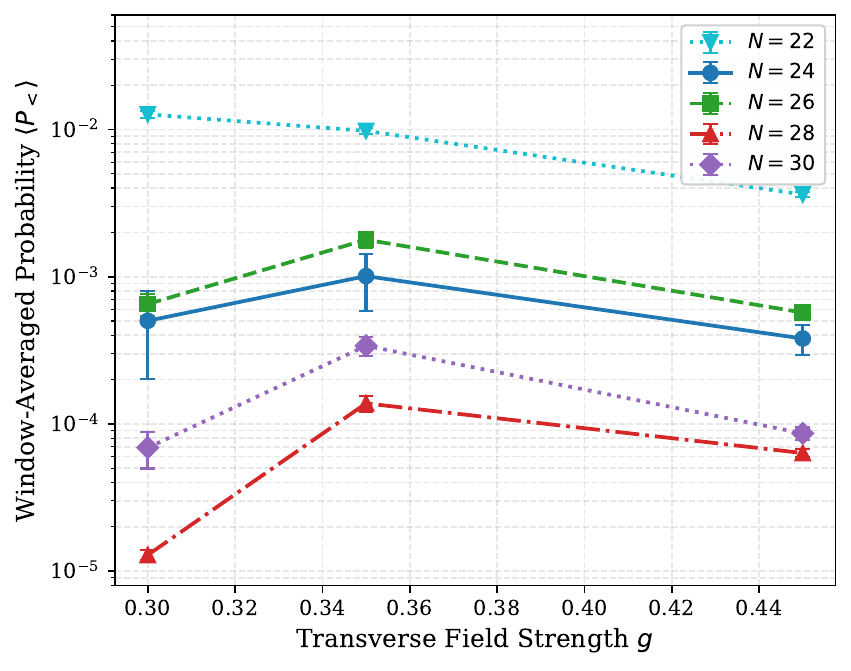}
        \caption{}
        \label{fig:pless_dh_g_a}
    \end{subfigure}\hfill
    \begin{subfigure}[b]{0.48\textwidth}
        \centering
        \includegraphics[width=\textwidth]{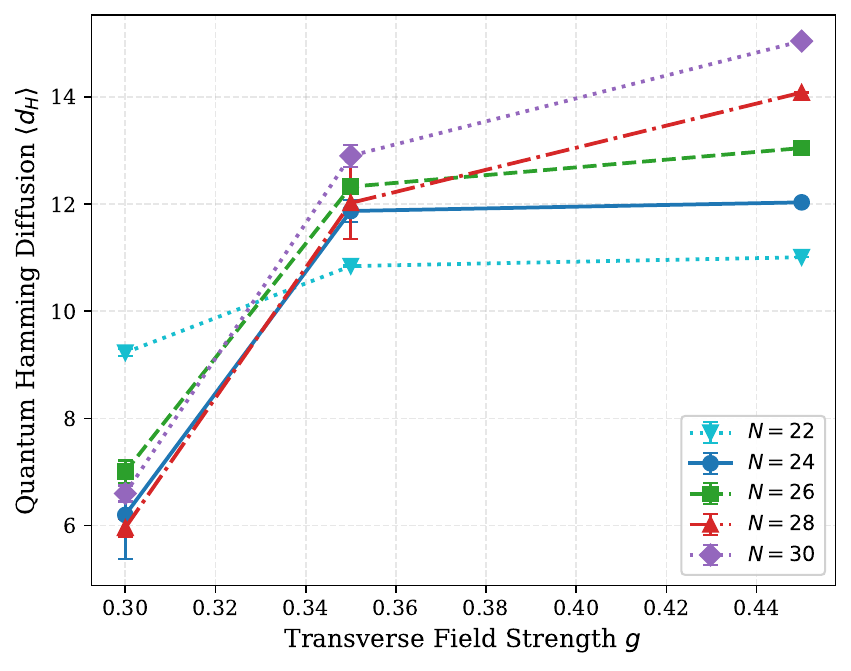}
        \caption{}
        \label{fig:pless_dh_g_b}
    \end{subfigure}
    \caption{Effect of transverse field $g$ on quantum escape and Hamming-space spreading for system sizes $N \in \{22, 24, 26, 28, 30\}$.
    (a)~Window-averaged escape probability $\langle P_< \rangle$ versus $g$, showing the field dependence of the low-energy yield within the chosen window.
    (b)~Mean Hamming displacement $\langle d_H \rangle$ versus $g$, showing increasing displacement towards $N/2$. Error bars indicate standard error of the mean (SEM).}
    \label{fig:pless_dh_g}
\end{figure*}

Figure~\ref{fig:rise_rate_dmin} makes this connection through the
initial rise rate $\left.dP_</dt\right|_{\mathrm{rise}}$ and the
window-averaged probability $\langle P_<\rangle$ across the evaluated
traps, whose minimal barrier distances span $d_{\min}\in[2,13]$.
Comparing these two observables shows how the distance to a target and
the number of available targets enter at different stages of escape.

For a fixed barrier distance (for instance, $d_{\min} = 6$), window escape probabilities vary by nearly two orders of magnitude (ranging from $\langle P_< \rangle \approx 3.2 \times 10^{-4}$ up to $1.4 \times 10^{-2}$) depending on the trap's energy tier $E_0$. The deepest classical traps (Level~1, $E_0^*$, with barriers $d_{\min}=4$--$13$) systematically exhibit lower window escape probabilities ($\langle P_< \rangle \sim 2.7 \times 10^{-5} - 1.7 \times 10^{-3}$ across $N \in \{20, 22, 24, 26, 28, 30\}$), while excited local minima (Level~2, with $d_{\min}=2$--$11$) achieve higher escape probabilities (up to $\sim 10^{-2}$). This separation reflects the volume of the target set $\mathcal{S}_<$ defined before Eq.~\eqref{eq:target_counts}. For excited traps, $\mathcal{S}_<$ encompasses extensive lower-lying basins, whereas for Level~1 traps it is restricted to rare states near the global optimum ($P_{\mathrm{blind}} = 1.49 \times 10^{-8}$ at $N = 30$).

To understand the early part of this evolution, we first use the
distance in Eq.~\eqref{eq:d_min} to determine how escape begins. For a
continuous-time quantum walk, the shortest paths determine the leading
power of the transition probability~\cite{szigeti2019}. Here the
hypercube structure also fixes its coefficient. Consider a
target $s\in\mathcal{S}_<$ at distance $d=d_H(s,s_0)$, and let $m$ be a
nonnegative integer power of the Hamiltonian. Each off-diagonal factor
in Eq.~\eqref{eq:transverse_hamiltonian} flips one spin, so reaching $s$
requires at least $d$ such factors. At that order, the required flips can
occur in $d!$ different sequences, giving
    \begin{align}
        \langle s|\hat{H}(g)^m|s_0\rangle &= 0 \quad (m<d),\nonumber\\
        \langle s|\hat{H}(g)^d|s_0\rangle &= (-g)^d d!.
        \label{eq:shortest_path_matrix_element}
    \end{align}
The factorial cancels the Taylor coefficient of the evolution operator.
Since the Hamiltonian is real in the computational basis, transition
probabilities are even functions of time. Summing over the nearest
targets in Eq.~\eqref{eq:escape_probability} gives the exact onset law
    \begin{equation}
        P_<(t)=M_{d_{\min}}(gt)^{2d_{\min}}
        +\mathcal{O}(t^{2d_{\min}+2}).
        \label{eq:short_time_escape}
    \end{equation}
Distance fixes the power of time, while the number of nearest targets
fixes its coefficient. This provides a direct explanation of the
initial suppression in Fig.~\ref{fig:hardest_dynamics}.

The leading onset depends only on the shortest paths and their number.
To see where the landscape energies enter, we next examine the first
correction. For a fixed target at distance $d$, let $v_j$ be the configuration after
    $j$ flips along a shortest sequence, with $v_0=s_0$ and $v_d=s$.
Denote the full sequence by $\pi=(v_0,v_1,\ldots,v_d)$. Choose these $d!$
sequences with equal weight. Define the energy offset $\epsilon_j$
at $v_j$ using the normalization of Eq.~\eqref{eq:transverse_hamiltonian},
and denote their sum along the sequence by $S_\pi$,
    \begin{equation}
        \epsilon_j=\frac{E(v_j)-E_0}{N},\qquad
        S_\pi=\sum_{j=0}^{d}\epsilon_j.
        \label{eq:path_energy_offsets}
    \end{equation}
To keep track of the phase accumulated during a sequence, let $u_j$ be
the fraction of the total time spent at $v_j$. Integrating over the
ordered flip times gives a uniform distribution over $u_j\geq0$ with
$\sum_{j=0}^{d}u_j=1$. With these fractions, the energy averaged along
the path is $Y_s=\sum_{j=0}^{d}u_j\epsilon_j$. We characterize its
variation over both the paths and the time fractions by the variance
$V_s$. Writing $\mathbb{E}_\pi$ for the uniform average over paths gives
    \begin{align}
        V_s={}&\frac{\mathbb{E}_\pi\!\left[S_\pi^2+
                      \sum_{j=0}^{d}\epsilon_j^2\right]}{(d+1)(d+2)}
                \nonumber\\
             &-\left(\frac{\mathbb{E}_\pi S_\pi}{d+1}\right)^2\geq0.
        \label{eq:path_energy_variance}
    \end{align}
To see how this variance enters the transition probability, write
$\mathbb{E}_{\pi,u}$ for the average over paths and time fractions
$(u_0,\ldots,u_d)$. After removing the common phase $e^{-iE_0t/N}$,
the contribution with $d$ flips becomes
$(igt)^d\mathbb{E}_{\pi,u}[e^{-itY_s}]$.
Its squared modulus supplies the correction $-V_st^2$. In addition,
paths with two further flips contribute the correction already present
for the free transverse mixer. Combining the two contributions yields
    \begin{align}
        &|\langle s|e^{-it\hat{H}(g)}|s_0\rangle|^2\nonumber\\
        &\quad=(gt)^{2d}\Bigl[1-t^2\bigl\{
             g^2(N-2d/3)+V_s\bigr\}\nonumber\\
        &\hspace{35mm}+\mathcal{O}(t^4)\Bigr].
        \label{eq:short_time_energy_correction}
    \end{align}
The coefficient proportional to $g^2$ follows by expanding the free
mixer probability $\sin^{2d}(gt)\allowbreak\cos^{2(N-d)}(gt)$.
Thus the geometry determines the leading onset, and the spread of
energies along the paths determines its first landscape-dependent
correction.
{}{The nonnegative variance $V_s$ makes this first landscape
correction suppressive. The enhanced escape measured in the sampling
window therefore develops beyond the earliest shortest-path contribution,
as further spin flips and relative phases enter the evolution.}
This later evolution connects the window probability in
Eq.~\eqref{eq:window_average} and the first-peak growth indicator to the
field dependence observed in Fig.~\ref{fig:rise_rate_dmin}.
    
We can now compare this early-time structure with the transfer observed
inside the sampling window.
Broad Hamming-space spreading ($\langle d_H\rangle\approx N/2$,
Sec.~\ref{subsec:field_regimes}) occurs together with enhanced weight
on lower-energy configurations. This is the combination expected in
energy-selective population transfer~\cite{smelyanskiy2020,baldwin2018}.
The transverse field mixes configurations across Hamming shells, while
the phases generated by $\hat{H}_B/N$ give the transfer its energy
dependence. Window probabilities reach order $10^{-3}$ in the deepest
tier and $10^{-2}$ in the second tier, even among traps with
$d_{\min}=6$--$13$, where the number of possible kicks
$\binom{N}{d_{\min}}$ reaches order $10^7$.

Figure~\ref{fig:energy_hist} shows this directly for the deepest $N = 28$ trap.
At $t = 30$ the weight on the ground-state energy $E^{*} = 50$ is
    $2.6 \times 10^{-4}$, about $10^{4}$ times the blind fraction of
    $3.0 \times 10^{-8}$, while the mean energy $\langle E \rangle \approx 156$ stays
far below the random-string mean $N(N-1)/2 = 378$. The state therefore
combines broad Hamming displacement with a strong preference for low energies.
{}{The enhanced ground-state weight occurs while the mean
cost energy rises from the trap value $58$ to about $156$. Escape is
therefore accompanied by a redistribution of probability towards both
lower and higher classical energies.}

To quantify the preference for low energies within the distance shells,
we use Eq.~\eqref{eq:isotropic_bound}.
For the hardest orbits in Table~\ref{tab:hardest_seeds}, the target counts
are $M=8$ at $N=28$ and $M=16$ at $N=30$, with $d_{\min}=10$ and $13$,
respectively. The corresponding isotropic-kick bounds are
    $B_{\mathrm{iso}}=6.10\times10^{-7}$ and $1.34\times10^{-7}$.
The window probabilities exceed these bounds by approximately
    $293$ and $203$, respectively. Since the bound holds for every choice
of kick radius and for their mixtures, the enhancement requires a
preference for successful configurations within the Hamming shells.
The same bound applies to a time-averaged distribution, so the comparison
applies directly to the sampling window used in the simulations.

This change in the mean cost energy has a neat interpretation through
energy conservation. The relation between the cost and driver
energies~\cite{callison2021} connects the mean energy in
Fig.~\ref{fig:energy_hist} to the transverse spin polarization.
Write $\langle\cdot\rangle_t$ for expectation in the state evolved from
$|s_0\rangle$. Initially $\langle\hat{X}_i\rangle_0=0$, so continuous-time
evolution under Eq.~\eqref{eq:transverse_hamiltonian} conserves the total
energy $\langle\hat{H}(g)\rangle_t=E_0/N$. Consequently,
\begin{align}
    \langle\hat{H}_B\rangle_t
        &=E_0+gN\left\langle\sum_{i=1}^{N}\hat{X}_i\right\rangle_t,
        \nonumber\\
    \bigl|\langle\hat{H}_B\rangle_t-E_0\bigr|&\leq |g|N^2,
    \label{eq:energy_balance}
\end{align}
where the inequality follows because the sum of $N$ Pauli-$X$ operators
has eigenvalues between $-N$ and $N$. Define the variance of the full
Hamiltonian by $(\Delta H)^2_t=\langle\hat{H}(g)^2\rangle_t
-\langle\hat{H}(g)\rangle_t^2$.
Acting once on $|s_0\rangle$, the driver produces $N$ orthogonal one-flip
states, each with amplitude $-g$. The conserved variance is therefore
\begin{equation}
    (\Delta H)^2_t=g^2N.
    \label{eq:conserved_energy_variance}
\end{equation}
{}{At fixed $N$ and $g$, every initial basis trap has the
same standard deviation $|g|\sqrt{N}$ of the total energy, even though
their escape probabilities differ. Their different yields probe the
geometry of the lower-energy targets and the energies encountered on
the way to them.}

The low mean cost energy suggests a further question. How much of the
enhanced ground-state weight follows from this mean alone, and how much
depends on the way probability moves through the landscape? To make this
comparison, write $P(E,t)$ for the probability of measuring classical
energy $E$ at time $t$, and $\bar E(t)=\sum_E E P(E,t)$ for its mean.
Let $\mathcal{N}(E)$ count all configurations of energy $E$, including
their symmetry partners. Among distributions over these configurations
with a prescribed mean, maximizing the Shannon entropy gives a canonical
cost distribution~\cite{jaynes1957}. Denote its dimensionless inverse
temperature by $\beta$ and its partition function by
$Z_{\mathrm{can}}(\beta)$. With the energy normalization of
Eq.~\eqref{eq:transverse_hamiltonian}, they enter as
\begin{align}
 P_{\mathrm{can}}(E;\beta)
     &=\frac{\mathcal{N}(E)e^{-\beta E/N}}{Z_{\mathrm{can}}(\beta)},
     \nonumber\\
 Z_{\mathrm{can}}(\beta)
     &=\sum_E\mathcal{N}(E)e^{-\beta E/N}.
 \label{eq:canonical_cost_reference}
\end{align}
Thus the density of states fixes the relative abundance of each energy
level, while the Boltzmann factor fixes the preference for low energies.
For each snapshot, define the effective inverse temperature
$\beta_{\mathrm{eff}}(t)$ by matching its mean,
\begin{equation}
 \sum_E E P_{\mathrm{can}}\bigl(E;\beta_{\mathrm{eff}}(t)\bigr)
       =\bar E(t).
 \label{eq:effective_temperature_match}
\end{equation}
The canonical mean decreases monotonically with $\beta$, since its
derivative is minus the canonical energy variance divided by $N$.
Consequently, this condition determines a unique inverse temperature,
after which every energy-level probability is fixed.

\begin{figure*}[t]
\centering
\begin{subfigure}[b]{0.49\textwidth}
    \centering
    \includegraphics[width=\textwidth]{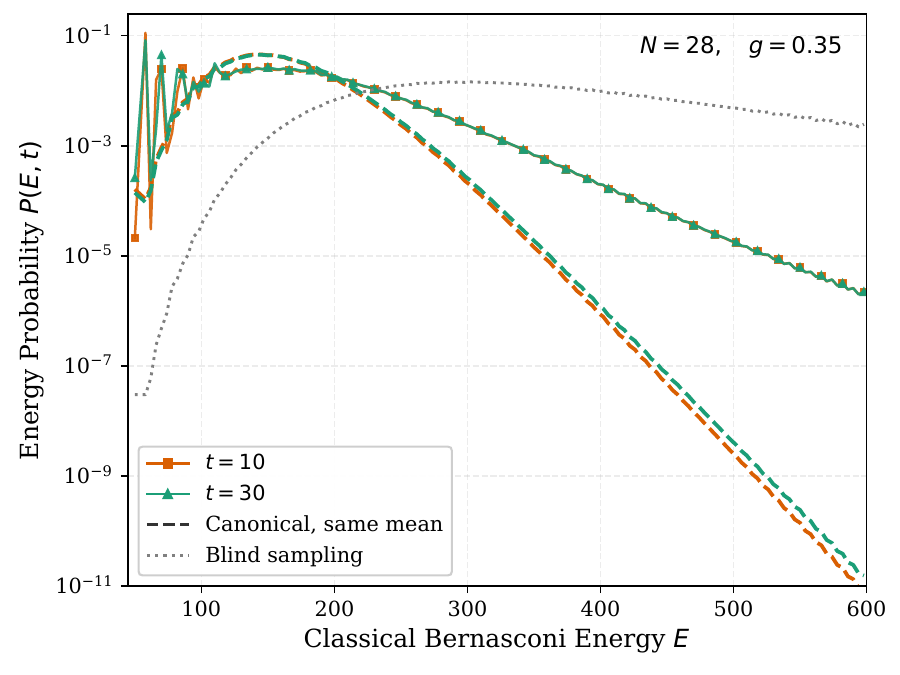}
    \caption{}
    \label{fig:thermal_test_a}
\end{subfigure}\hfill
\begin{subfigure}[b]{0.49\textwidth}
    \centering
    \includegraphics[width=\textwidth]{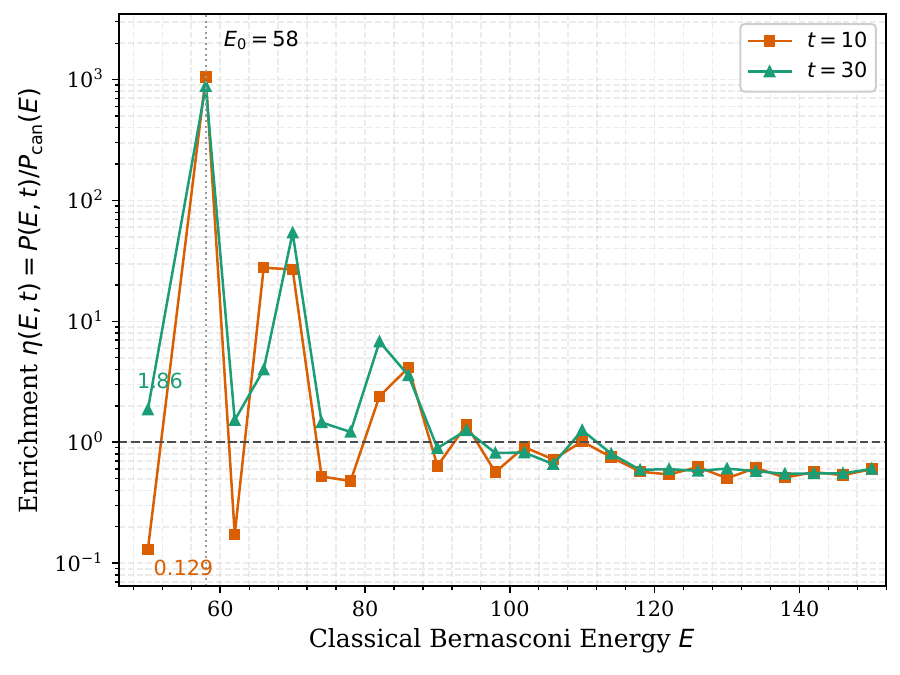}
    \caption{}
    \label{fig:thermal_test_b}
\end{subfigure}
\caption{Canonical cost comparison for the deepest $N=28$ trap
($E_0=58$, $g=0.35$), using the archived snapshots in
Fig.~\ref{fig:energy_hist}.
(a)~Recorded probabilities at $t=10$ and $30$ (solid lines with
markers) and canonical distributions with the same recorded first
moments (dashed lines in matching colours). The grey dotted curve
is $\mathcal{N}(E)/2^N$. Each canonical curve is normalized using
the complete density of states, including energies beyond the displayed
range. (b)~Enrichment $\eta(E,t)$ from
Eq.~\eqref{eq:canonical_enrichment} in the low-energy region.
The horizontal line denotes agreement with the canonical prediction.
The ground-state enrichment rises from $0.129$ to $1.86$, while the
initial trap energy retains a much larger excess. Lines join discrete
energy levels to guide the eye.}
\label{fig:thermal_test}
\end{figure*}

For the $N=28$ run in Fig.~\ref{fig:energy_hist}, the saved density of
states contains $803$ distinct energies and sums to exactly $2^{28}$
configurations. We use the recorded first moments $154.0269$ at $t=10$
and $155.8906$ at $t=30$ in
Eq.~\eqref{eq:effective_temperature_match}.\footnote{The saved
histograms retain energies $E<600$ and contain more than $0.9999766$
of the total probability. The quoted first moments use this retained
weight without renormalization. Since the largest Bernasconi energy
at $N=28$ is $\sum_{k=1}^{27}k^2=6930$, the omitted weight can increase
either mean by at most $0.163$. Repeating the fit over this entire
interval gives a ground-state enrichment between $1.86$ and $1.89$
at $t=30$.}
The corresponding inverse temperatures are $1.4351$ and $1.3968$.
There are eight ground-state configurations at $E^*=50$, so
Eq.~\eqref{eq:canonical_cost_reference} predicts ground-state probabilities
of $1.61\times10^{-4}$ and $1.40\times10^{-4}$, respectively.
The measured values are $2.09\times10^{-5}$ and $2.60\times10^{-4}$.
To compare the distributions at every energy, define the enrichment
$\eta(E,t)$ as the measured probability relative to this canonical
prediction,
\begin{equation}
 \eta(E,t)=\frac{P(E,t)}
 {P_{\mathrm{can}}\bigl(E;\beta_{\mathrm{eff}}(t)\bigr)}.
 \label{eq:canonical_enrichment}
\end{equation}
Figure~\ref{fig:thermal_test} shows both distributions and this ratio.
At the ground-state energy, $\eta(E^*,10)=0.129$ and
$\eta(E^*,30)=1.86$. In addition, the probability at the initial
energy $E_0=58$ remains $0.0826$ at $t=30$, about $881$ times
its canonical value. Substantial excess weight also persists in the
high-energy tail.
{}{Between these two snapshots, the ground-state
probability grows by a factor of $12.5$ while the mean cost energy
changes by only about $1.2\%$. The transfer therefore carries
information beyond the mean, simultaneously retaining weight at
the initial trap energy and increasing the population of the lower
ground-state level.}

The distinction between cost energy and total energy is useful when
interpreting this test. For a canonical reference based on the full
quantum Hamiltonian, let $\beta_{\mathrm q}$ be its inverse temperature
and let $\hat\Pi_E=\sum_{s:E(s)=E}|s\rangle\langle s|$ project onto
the configurations of classical energy $E$. Taking the trace over the
full spin Hilbert space gives its prediction for a cost measurement,
\begin{equation}
 P_{\mathrm q}(E;\beta_{\mathrm q})=
 \frac{\operatorname{Tr}\!\left[\hat\Pi_E
             e^{-\beta_{\mathrm q}\hat H(g)}\right]}
      {\operatorname{Tr}e^{-\beta_{\mathrm q}\hat H(g)}}.
 \label{eq:quantum_canonical_reference}
\end{equation}
Here $\beta_{\mathrm q}$ is fixed by the conserved mean total energy
$E_0/N$, and the transverse field remains inside the Gibbs operator.
Equilibration within conserved symmetry sectors also retains their
initial weights~\cite{dalessio2016}.
Equation~\eqref{eq:canonical_enrichment} quantifies the additional
energy selection beyond a mean-matched
canonical cost distribution. Field and time enter that reference
through the measured mean, while $\eta$ records the further structure
generated by the dynamics.

Energy conservation also constrains how broadly the classical energy can
be sampled. A uniform distribution over bitstrings has mean energy
$N(N-1)/2$, as follows by averaging Eq.~\eqref{eq:expanded_energy}.
For a seed below this mean, define $g_{\mathrm u}(N,E_0)$ as the
lower bound on the field magnitude required to reach the uniform
mean. Equation~\eqref{eq:energy_balance} gives
\begin{equation}
    |g|\geq g_{\mathrm u}(N,E_0)
       \equiv\frac{N(N-1)/2-E_0}{N^2}.
    \label{eq:uniform_energy_condition}
\end{equation}
Table~\ref{tab:hardest_seeds} lists $g_{\mathrm u}$ for the six
hardest orbits. Its values lie between $0.3802$ and $0.4089$.
For the $N=28$, $E_0=58$ seed, $g_{\mathrm u}=0.40816$.
At the adopted field $g=0.35$, Eq.~\eqref{eq:energy_balance} gives
$\langle\hat{H}_B\rangle_t\leq332.4$, below the uniform mean $378$.
The continuous-time dynamics thus retains a mean energy below the
uniform-string value even as the state spreads in Hamming distance.
At $g=0.45$, the field exceeds $g_{\mathrm u}$ for all six seeds,
so this energy constraint allows the mean to reach the uniform value.
The actual distribution is then determined by the transfer dynamics.
These conservation relations refer to the continuous-time Hamiltonian used
as the reference for the Trotter evolution in Sec.~\ref{sec:methodology}.
\subsection{Effect of the transverse field}
\label{subsec:field_regimes}
The preceding analysis shows how the landscape shapes the transfer.
We now ask how this transfer changes when the transverse field becomes
stronger. The field $g$ controls the balance between the diagonal
Bernasconi potential $\hat{H}_B/N$ and the spin flips generated by
$\sum_i\hat{X}_i$. Figure~\ref{fig:pless_dh_g} follows this balance
through the window-averaged escape probability $\langle P_<\rangle$
and mean Hamming displacement $\langle d_H\rangle$, for
$g\in\{0.30,0.35,0.45\}$ and $N\in\{22,24,26,28,30\}$.

The transverse field alone gives the continuous-time walk on the hypercube~\cite{moore2002,mulken2011}, providing a useful reference for this spreading.
With the diagonal term removed, Eq.~\eqref{eq:mixer_factorization} makes
each spin flip independently with probability $q(t)=\sin^2(gt)$ upon
measurement. Denote the resulting escape probability by
$P_{\mathrm{free}}(t)$ and the mean Hamming displacement by
$\langle d_H(t)\rangle_{\mathrm{free}}$. Then
\begin{align}
P_{\mathrm{free}}(t)
        &=\sum_{d=0}^{N}M_d\,q(t)^d[1-q(t)]^{N-d},\nonumber\\
    \langle d_H(t)\rangle_{\mathrm{free}}&=Nq(t).
    \label{eq:free_mixer_probabilities}
\end{align}
{}{For the free mixer, the field enters only through
$gt$, so changing $g$ rescales the time along the same evolution.
Once the Bernasconi term is present, changing $g$ also changes the balance
between spin flips and the phases accumulated from the fixed energy landscape.}

To bound the escape produced by this free spreading, consider a flip
probability $q$ and a target at distance $d$. The probability
$q^d(1-q)^{N-d}$ is largest
at $q=d/N$. Over the allowed target distances
$d_{\min}\leq d\leq N-d_{\min}$, this maximum is largest at the endpoints.
Summing the bound over the $M$ targets gives
\begin{equation}
P_{\mathrm{free}}(t)\leq
M\left(\frac{d_{\min}}{N}\right)^{d_{\min}}
    \left(1-\frac{d_{\min}}{N}\right)^{N-d_{\min}}.
    \label{eq:free_mixer_bound}
\end{equation}
The bound holds for every time and field, and hence for every sampling
window. For the hardest $N=28$ and $N=30$ seeds it gives
$9.50\times10^{-8}$ and $1.95\times10^{-8}$, respectively.
The observed window probabilities are about $1.88\times10^3$ and
$1.39\times10^3$ times larger. The diagonal Bernasconi term is therefore
essential to the observed transfer into lower-energy configurations,
while the mixer alone already produces a large Hamming displacement.

With this reference in place, the field dependence in
Fig.~\ref{fig:pless_dh_g} traces a progression
from partial spreading to broad exploration. At $g=0.30$, the mean
Hamming displacement is approximately $5.96$--$9.23$ for
$N\in\{22,24,26,28,30\}$, or $43\%$--$84\%$ of $N/2$.
For example, $\langle d_H\rangle=6.60\pm0.15$ at $N=30$, whose
hypercube midpoint is $15$. At $N=20$ the displacement is $7.15$,
or $71\%$ of $N/2$. The potential retains substantial weight near the
initial trap. This slower spreading can nevertheless produce a larger
late-time escape probability for excited traps, as discussed in
Sec.~\ref{subsec:coupling_crossover}.

Increasing the field to $g=0.35$ produces a pronounced expansion.
For $N\in\{22,24,26,28,30\}$, the mean displacement rises to
$10.84$--$12.90$, or $86\%$--$99\%$ of $N/2$, reaching
$12.90\pm0.20$ at $N=30$. The wavepacket now explores configurations
well beyond the initial basin while retaining elevated escape
probabilities, with the averages in Fig.~\ref{fig:pless_dh_g} ranging
from about $1.38\times10^{-4}$ to $9.79\times10^{-3}$.

Increasing the field further to $g=0.45$ brings the mean displacement
to approximately $N/2$.
Its values are $11.01$, $12.03$, $13.05$, $14.08$ and $15.04$ for
$N=22$, $24$, $26$, $28$ and $30$, respectively. However, the escape
probability decreases at the same time, reaching about
$8.63\times10^{-5}$ at $N=30$. Stronger driving thus gives a larger
displacement but a smaller yield of lower-energy configurations.
This decrease occurs above the energy-conservation bound
$g_{\mathrm u}$ listed in Table~\ref{tab:hardest_seeds}, where
a wider range of mean cost energies becomes accessible.
The comparison with Eq.~\eqref{eq:free_mixer_bound} shows why both
observables are useful. Efficient escape requires the diagonal
potential to shape the distribution as the transverse field spreads it.
{}{A superposition with equal weight on a configuration
and its globally inverted partner has mean Hamming displacement $N/2$
from the initial trap. Together with this displacement, the energy
distribution identifies which parts of the landscape carry the probability.}

The rapid change from partial spreading at $g=0.30$ to displacement
above $85\%$ of $N/2$ at $g=0.35$ marks a finite-time crossover to
broad Hamming-space spreading at the sizes studied. The adopted
field, $g=0.35$, combines this broad spreading with a strong preference
for low-energy configurations within the chosen window. A similar competition between
spreading and energy selectivity appears in population transfer in
random-energy and impurity-band models~\cite{baldwin2018,smelyanskiy2020,parolini2020},
whose extended-state regimes have also been studied
analytically~\cite{faoro2019}. Here the deterministic Bernasconi
potential retains the energetic preference while the field supplies
enough mixing to traverse wide Hamming barriers. At weaker fields the
transfer builds more slowly, and at stronger fields the additional
spreading reduces the low-energy yield within the same window.

\subsection{Growth of escape over time}
\label{subsec:coupling_crossover}
\label{subsec:window}
\begin{figure}[t]
    \centering
    \includegraphics[width=\columnwidth]{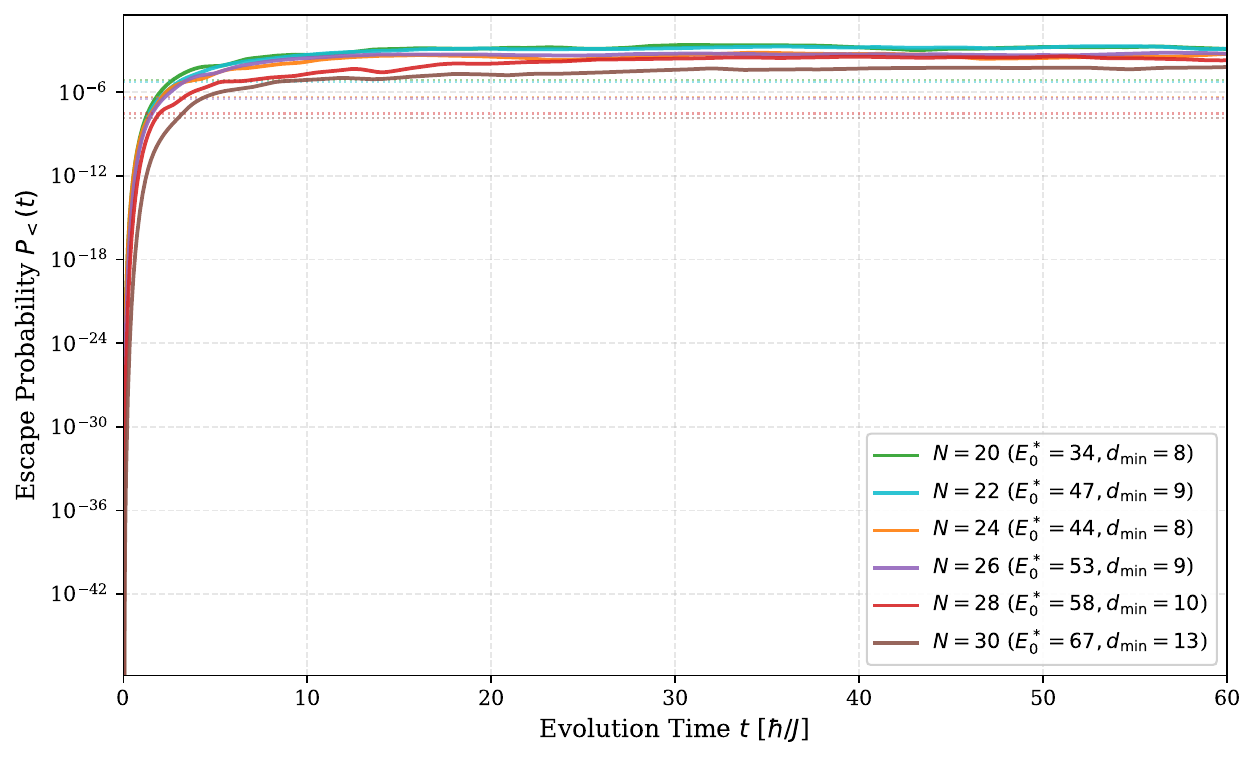}
    \caption{Continuous-time escape probability $P_<(t)$ for the hardest local minimum at each system size $N \in \{20, 22, 24, 26, 28, 30\}$ at $g = 0.35$. Horizontal dashed lines indicate blind search probability $P_{\mathrm{blind}}$.}
    \label{fig:hardest_dynamics}
\end{figure}

The field scan compares averages over one interval. To see how those
averages develop, we now follow the full escape trajectories and then
vary the time at which they are compared. Figure~\ref{fig:hardest_dynamics}
shows $P_<(t)$ for the hardest local minimum at each
$N\in\{20,22,24,26,28,30\}$, using $g=0.35$.

As expected from the short-time expansion, the trajectories begin with
a delay of order $t\lesssim3$--$5$.
All these seeds have $d_{\min}\geq6$, so reaching the target requires
several spin flips and $P_<(1)\lesssim10^{-7}$ across the sizes
shown. This suppression follows from the power of time in
Eq.~\eqref{eq:short_time_escape}. During $t\approx5$--$30$, the
escape probability grows by two to four orders of magnitude as the
state develops weight in lower-energy basins, reaching order $10^{-3}$
for the larger-probability trajectories.

For the deepest Level~1 traps at $g = 0.35$ the late-time behaviour differs by instance. At $N = 28$ ($E_0 = 58$, $d_{\min} = 10$) the probability has essentially plateaued by the end of the window ($P_<(60) = 1.96 \times 10^{-4}$ against a window average of $1.79 \times 10^{-4}$), whereas at $N = 30$ ($E_0 = 67$) it is still rising at $t = 60$ ($P_<(60) = 1.43 \times 10^{-4}$ for the $d_{\min} = 11$ orbit, $2.5\times$ its window average). The late rise is more pronounced for Level~2 traps and for $g=0.30$,
as the comparisons in Table~\ref{tab:window_underestimation} will show. For these rising trajectories, the probability measured at $t=60$ exceeds the fixed-window average.

To appreciate the size of this transfer, we compare these trajectories
with the fraction of configurations that qualify as targets.
For the hardest orbit at each size, Table~\ref{tab:hardest_seeds} lists
the blind success probability $P_{\mathrm{blind}}=M/2^N$, using the
target count in Eq.~\eqref{eq:target_counts}. It falls by more than two
orders of magnitude, from $7.63\times10^{-6}$ at $N=20$ to
$1.49\times10^{-8}$ at $N=30$. The quantum window probabilities range
from $1.67\times10^{-3}$ to $2.71\times10^{-5}$, showing a strong
concentration of weight on the sparse lower-energy set.

Classical local search also uses the landscape to improve its proposals,
so it gives a more demanding comparison than the blind fraction. Their best rate for the hardest orbit is
$4.79\times10^{-5}$ at $N=20$ and $3.34\times10^{-6}$ at $N=30$,
where the barrier distances in the table grow from $8$ to $13$.
Relative to this best baseline, quantum sampling gives
$R=8.1$--$35.2$ for the hardest individual orbits within the fixed
window $[10.0,38.5]$. Table~\ref{tab:ratio_summary} gives the
corresponding medians over all orbits in each tier.

\begin{table*}[t]
    \centering
    \small
    \caption{Hardest single orbit at each system size ($g = 0.35$), defined as the Level~1 orbit with the largest $d_{\min}$ (first evaluated orbit on ties). $P_{\mathrm{blind}}$ is the fraction of all configurations below $E_0$. The ratio $R = \langle P_< \rangle / r_{\mathrm{classical}}^*$ uses the best classical method listed. $g_{\mathrm u}$ is the necessary field bound for reaching the uniform mean cost energy in Eq.~\eqref{eq:uniform_energy_condition}. These are single-orbit values. Table~\ref{tab:ratio_summary} and the abstract quote medians over all orbits in a tier.}
    \label{tab:hardest_seeds}
    \resizebox{\textwidth}{!}{%
    \begin{tabular}{ccccccccc}
        \toprule
        $N$ & $E_0$ & $g_{\mathrm u}$ & $d_{\min}$ & $P_{\mathrm{blind}}$ & Window $\langle P_< \rangle$ & Best Classical Method & $r_{\mathrm{classical}}^*$ & Ratio $R$ \\
        \midrule
        20 & 34 & $0.3900$ &  8 & $7.63 \times 10^{-6}$ & $1.67 \times 10^{-3}$ & Tabu Direct & $4.79 \times 10^{-5}$ & $\mathbf{34.9\times}$ \\
        22 & 47 & $0.3802$ &  9 & $5.72 \times 10^{-6}$ & $1.33 \times 10^{-3}$ & Kick-SD ($d=12$) & $8.52 \times 10^{-5}$ & $\mathbf{15.6\times}$ \\
        24 & 44 & $0.4028$ &  8 & $4.77 \times 10^{-7}$ & $4.13 \times 10^{-4}$ & Tabu Restart & $2.05 \times 10^{-5}$ & $\mathbf{20.2\times}$ \\
        26 & 53 & $0.4024$ &  9 & $3.58 \times 10^{-7}$ & $5.02 \times 10^{-4}$ & Tabu Kick ($d=10$) & $2.38 \times 10^{-5}$ & $\mathbf{21.1\times}$ \\
        28 & 58 & $0.4082$ & 10 & $2.98 \times 10^{-8}$ & $1.79 \times 10^{-4}$ & Tabu Kick ($d=4$) & $5.07 \times 10^{-6}$ & $\mathbf{35.2\times}$ \\
        30 & 67 & $0.4089$ & 13 & $1.49 \times 10^{-8}$ & $2.71 \times 10^{-5}$ & Tabu Kick ($d=12$) & $3.34 \times 10^{-6}$ & $\mathbf{8.1\times}$ \\
        \bottomrule
    \end{tabular}%
    }
\end{table*}

The same trajectories also let us revisit the choice of field. Denote
by $g^*$ the transverse field giving the largest escape probability
at the observation time. Although $g=0.35$ produces faster early
transfer, the initially slower $g=0.30$ trajectory can continue to grow
until it matches or exceeds the stronger-field result. Thus the
preferred field depends on how long the move is allowed to develop.

\begin{table*}[t]
\centering
\small
\caption{Escape probability $P_<(t)$ comparing couplings $g = 0.30$ and $g = 0.35$ for evaluated traps across $N \in \{22, 26, 30\}$. Early on ($t \le 10$), $g = 0.35$ leads. Later ($t \ge 45 - 50$), $g = 0.30$ maintains amplitude in low-energy basins, overtaking $g = 0.35$. In the probability columns, the argument denotes the field strength at the listed time.}
\label{tab:coupling_crossover}
\resizebox{\textwidth}{!}{%
\begin{tabular}{lccccc}
\toprule
$N$ & Time $t$ & $P_<(0.30)$ & $P_<(0.35)$ & $\frac{P_<(0.30)}{P_<(0.35)}$ & Dynamic Regime \\
\midrule
\textbf{$N = 22$ ($E_0 = 47, d_{\min} = 6$)} & $5.0$ & $3.22 \times 10^{-4}$ & $7.45 \times 10^{-4}$ & $0.43$ & $g = 0.35$ faster rise \\
& $10.0$ & $5.76 \times 10^{-4}$ & $1.34 \times 10^{-3}$ & $0.43$ & $g = 0.35$ faster rise \\
& $38.5$ & $1.36 \times 10^{-3}$ & $1.66 \times 10^{-3}$ & $0.82$ & $g = 0.30$ catching up \\
& $50.0$ & $1.59 \times 10^{-3}$ & $8.07 \times 10^{-4}$ & $\mathbf{1.96}$ & \textbf{$g = 0.30$ superior} \\
& $60.0$ & $1.56 \times 10^{-3}$ & $1.14 \times 10^{-3}$ & $\mathbf{1.37}$ & \textbf{$g = 0.30$ superior} \\
\midrule
\textbf{$N = 26$ ($E_0 = 61, d_{\min} = 9$)} & $10.0$ & $1.07 \times 10^{-4}$ & $5.11 \times 10^{-4}$ & $0.21$ & $g = 0.35$ faster rise \\
& $38.5$ & $3.45 \times 10^{-3}$ & $4.94 \times 10^{-3}$ & $0.70$ & $g = 0.30$ catching up \\
& $50.0$ & $6.95 \times 10^{-3}$ & $4.49 \times 10^{-3}$ & $\mathbf{1.55}$ & \textbf{$g = 0.30$ superior} \\
& $60.0$ & $9.19 \times 10^{-3}$ & $6.11 \times 10^{-3}$ & $\mathbf{1.50}$ & \textbf{$g = 0.30$ superior} \\
\midrule
\textbf{$N = 30$ ($E_0 = 75, d_{\min} = 4$)} & $10.0$ & $9.16 \times 10^{-5}$ & $3.10 \times 10^{-4}$ & $0.30$ & $g = 0.35$ faster rise \\
& $38.5$ & $8.87 \times 10^{-4}$ & $9.99 \times 10^{-4}$ & $0.89$ & $g = 0.30$ catching up \\
& $50.0$ & $1.41 \times 10^{-3}$ & $1.07 \times 10^{-3}$ & $\mathbf{1.31}$ & \textbf{$g = 0.30$ superior} \\
& $60.0$ & $1.72 \times 10^{-3}$ & $1.26 \times 10^{-3}$ & $\mathbf{1.37}$ & \textbf{$g = 0.30$ superior} \\
\bottomrule
\end{tabular}%
}
\end{table*}

Table~\ref{tab:coupling_crossover} connects the preferred field to the
time available for the move. At $g=0.35$, the state spreads rapidly
beyond the initial basin and develops a larger early escape
probability. At $g=0.30$, the initial growth is slower, but the
probability continues to rise beyond $t=40$ and can overtake the
stronger-field trajectory. The same competition between exploration
and energetic preference therefore appears both in the field scan
and in the ordering of the late-time probabilities. The dependence
on observation time is also natural in quantum-walk optimization and
population transfer~\cite{callison2019,callison2021,parolini2020}.
{}{Under unitary evolution, probability can flow both into
and out of the lower-energy set. The crossing of trajectories at different
fields therefore makes the measurement time part of the escape move,
alongside the field strength.}

This time dependence also explains the role of the measurement
window. Fixing the window gives a common protocol for every instance,
while following the trajectories out to $t=60$ reveals how its choice
affects the measured yield. Several Level~2 orbits and
$g=0.30$ runs continue to rise, as does the $d_{\min}=11$ orbit at
$N=30$ under $g=0.35$. Other trajectories, including the deepest
$N=24$ and $N=28$ traps at $g=0.35$, have largely plateaued within
the observation interval. Table~\ref{tab:window_underestimation}
compares the fixed-window average with the observed endpoint
probability for representative traps.

\begin{table}[t]
\centering
\small
\caption{Comparison of window-averaged escape probability $\langle P_< \rangle_{[10, 38.5]}$ versus late-time escape probability $P_<(t = 60)$ for deep traps across $N \in \{24, 28, 30\}$.}
\label{tab:window_underestimation}
\resizebox{\columnwidth}{!}{%
\begin{tabular}{lcccccc}
\toprule
$N$ & Trap $E_0$ & $d_{\min}$ & Field $g$ & $\langle P_< \rangle_{[10, 38.5]}$ & $P_<(t = 60)$ & Factor \\
\midrule
24 & 44 &  8 & 0.30 & $7.59 \times 10^{-5}$ & $7.09 \times 10^{-5}$ & $0.93\times$ \\
24 & 48 &  9 & 0.30 & $9.28 \times 10^{-4}$ & $3.99 \times 10^{-3}$ & $4.30\times$ \\
28 & 58 & 10 & 0.35 & $1.79 \times 10^{-4}$ & $1.96 \times 10^{-4}$ & $1.10\times$ \\
28 & 62 & 11 & 0.30 & $1.37 \times 10^{-5}$ & $3.69 \times 10^{-5}$ & $2.69\times$ \\
30 & 67 & 11 & 0.35 & $5.65 \times 10^{-5}$ & $1.43 \times 10^{-4}$ & $2.54\times$ \\
30 & 67 & 13 & 0.35 & $2.71 \times 10^{-5}$ & $6.34 \times 10^{-5}$ & $2.34\times$ \\
30 & 67 & 13 & 0.30 & $1.89 \times 10^{-6}$ & $3.60 \times 10^{-6}$ & $1.90\times$ \\
\bottomrule
\end{tabular}%
}
\end{table}

For the rising trajectories in Table~\ref{tab:window_underestimation},
$P_<(60)$ exceeds the window average by factors up to $4.30$.
At $N=30$, $E_0=67$ and $d_{\min}=11$, for example, the
$g=0.35$ trajectory has
$\langle P_<\rangle_{[10,38.5]}=5.65\times10^{-5}$ and
$P_<(60)=1.43\times10^{-4}$, a factor of $2.54$.
For the hardest $d_{\min}=13$ orbit at the same field, the
fixed-window ratio is $R=8.1$ and the endpoint probability is
$2.34$ times the window average. These observations identify a
further gain in sample yield available at the measured later time.
They complement the $R>1$ medians at every size and tier under the
same field and sampling window.

The same effect becomes more pronounced at the weaker field.
At $N=30$, $E_0=67$ and $d_{\min}=13$, $g=0.30$ gives
$\langle P_<\rangle=1.89\times10^{-6}$ within $[10,38.5]$,
or $R=0.57$ relative to $r^*_{\mathrm{classical}}=3.34\times10^{-6}$.
By $t=60$, the escape probability reaches $3.60\times10^{-6}$,
a factor of $1.90$ higher and slightly above the classical rate.
The additional evolution time is therefore a physical resource that
can change the ranking of the escape moves.

\section{Conclusions}
\label{sec:conclusion}
The numerical comparison and analytical results together identify the
physical ingredients of successful escape. We bring these observations
together here and use them to outline the next steps for understanding
and implementing the dynamics.

Real-time transverse-field dynamics provides an effective escape move
from the deep local minima of the Bernasconi model. The GPU
statevector calculations cover $N=20$--$30$, up to approximately
$1.07\times10^9$ basis states, and begin from the complete set of
symmetry-distinct traps in the two lowest excited tiers. With the
same field and sampling window across all sizes, the median yield of
lower-energy configurations exceeds the best classical perturbation
rate by factors of $11.9$--$35.2$ for Level~1 and $13.4$--$27.2$
for Level~2. The improvement remains a constant factor across this
range of sizes, including the larger instances at $N=28$ and $30$.

The analytical results explain how this gain develops from the
structure of the dynamics.
The minimum target distance fixes the power of the short-time onset,
the number of nearest targets fixes its coefficient, and the
intermediate energies enter through the first correction. By the
sampling times, this initially constrained motion has developed into
broad Hamming-space spreading together with a strong
preference for low-energy configurations. For the hardest $N=28$ and
$N=30$ seeds, the window probabilities exceed the bound for every
isotropic kick distribution by factors of approximately $293$ and
$203$. The free-mixer bounds give an additional direct measure of the
role played by the Bernasconi potential. Energy conservation explains
how the mean cost energy remains below the uniform-string value
during this broad spreading.
{}{The canonical cost comparison further separates
the effect of this low mean energy from the selection generated by
the dynamics. At $N=28$, the ground-state weight rises from below
its mean-matched canonical value at $t=10$ to an enhancement of
about $1.9$ at $t=30$, while substantial weight persists at the
initial trap energy.}

The field scan brings the geometry and energy dependence together. At $g=0.30$, the
mean displacement remains at $43\%$--$84\%$ of $N/2$, falling to
at most $54\%$ for $N\geq24$. At $g=0.35$ it exceeds $85\%$
of $N/2$, while the low-energy yield remains enhanced. At $g=0.45$,
the displacement approaches $N/2$ and the escape probability falls.
The operating field thus lies just beyond the rapid increase in
spreading, where spatial exploration and energetic preference work
together. The later overtaking by $g=0.30$ shows that the optimum
also depends on the duration of the move. Even as the blind target
fraction falls to $1.49\times10^{-8}$ at $N=30$, the measured
window probabilities remain of order $10^{-5}$--$10^{-3}$ for the
hardest traps.

Turning from the escape mechanism to its implementation, the ratio
$R$ characterizes sample efficiency, while state preparation and
evolution determine the cost of generating each sample. Full-statevector
simulation requires resources exponential in $N$ and reaches about
$N=30$ on a single GPU in this study. The circuit estimate identifies
the two- and four-body phases as the main hardware cost. These resource
requirements give a concrete target for improved simulation and
compilation while the present unitary dynamics establishes the escape
mechanism.

These observations also suggest a natural next analytical step, namely
to connect the Bernasconi spectrum near
$E_0$ to the transfer rate and its field dependence, building on the
population-transfer treatments of
Refs.~\cite{smelyanskiy2020,baldwin2018}. The short-time expansion and
the bounds in Eqs.~\eqref{eq:isotropic_bound} and
\eqref{eq:free_mixer_bound} already constrain this development using
the existing trajectories. In addition, time-dependent fields $g(t)$ offer a way
to combine rapid initial spreading with the larger late-time yield
seen at weaker coupling. Matrix-product-state evolution and
variational time-evolution methods provide possible routes to larger
instances, with their cost set by the entanglement generated during
the escape~\cite{schollwock2011,haegeman2011,paeckel2019}.

Finally, sample-based diagonalization and Krylov methods help clarify
the role of the sampled states~\cite{robledo2025,yu2025}. For the
diagonal cost Hamiltonian used here, its restriction to a span of sampled
computational-basis states has those sampled energies as its eigenvalues.
Its minimum is therefore the best energy already sampled, which also
remained unchanged when descent was applied to the lowest-energy samples
in the present runs. The quantum contribution lies in the distribution
that makes these low-energy configurations available to the classical search.

\section*{Acknowledgements}
The authors thank the Sharanga HPC cluster for providing the computational facility.

RB thanks support from an intramural grant CDRF C2/24/282 from BITS Pilani, and from ANRF India through ANRF/ARG/2025/009241/PS. RB acknowledges stimulating discussions with Ritajit Majumdar at IBM India and Indrakshi Raychowdhury on quantum algorithms, which motivated the present work.

ChatGPT Astra (OpenAI) was used to assist in framing the logical flow
of the manuscript, and Gemini~3.8~Flash (Google) was used to assist in writing
and maintaining the simulation and analysis code. The authors take
responsibility for the final results and conclusions.

The code and numerical data are hosted on GitHub at
\href{https://github.com/ssk1001/bernasconi-krylov}{\nolinkurl{github.com/ssk1001/bernasconi-krylov}},
with access provided upon request to the authors.


\end{document}